\documentclass{aastex}
\usepackage{emulateapj5}
\usepackage{onecolfloat5}
\usepackage{apjfonts}

\input epsf

\let\gsim=\gtrsim

\makeatletter

\newcommand{\bfm}[1]{\mbox{${\bf #1}$}}

\newcommand{\skaco}[1]{\langle{#1}\rangle}

\lefthead{Takada $\&$ Jain}
\righthead{3-point Correlation Function of Spin-2 Field}

\begin{document}

\twocolumn[

\title{The three-point correlation function for spin-2 fields}

\author{Masahiro Takada
and Bhuvnesh Jain}
\affil{Department of Physics, University of Pennsylvania, 209 S. 33rd Street,
Philadelphia, PA 19104;  mtakada,bjain@hep.upenn.edu}

\begin{abstract}
The three-point correlation function (3PCF) of the spin-2 fields, 
cosmic shear and microwave background polarization, is a
statistical measure of non-Gaussian signals. At each vertex of
a triangle, the shear field has two independent components. 
The resulting eight possible 3PCFs were recently investigated by 
Schneider \& Lombardi (2002) and Zaldarriaga \& Scoccimarro (2002). 
Using rotation and parity transformations they posed the question: 
how many components of the shear 3PCF are non-zero and useful? 
We address this question using an analytical model and measurements 
from ray-tracing simulations. We show that all the eight 3PCFs are generally 
non-zero and have comparable amplitude. 
These eight 3PCFs can be used to improve the signal-to-noise from survey 
data and their configuration dependence can be used 
to separate the contribution from $E$- and $B$-modes. This separation 
provides a new and precise probe of systematic errors. 
We estimate the signal-to-noise 
for measuring the shear 3PCF from weak lensing surveys using
simulated maps that include the noise due to intrinsic ellipticities.
A deep lensing survey with area of order 10 square degrees would allow 
for the detection of the shear 3PCFs; a survey with area exceeding 100 
square degrees is needed for accurate measurements. 
\end{abstract}

]

\section{Introduction}

Can the three-point correlation function (3PCF) of a spin-2 field be a
useful probe of non-Gaussian signals in cosmology?  Most 
previous work has focused on the 3PCF or its Fourier transform, the
bispectrum, of scalar quantities such as the  galaxy number density
and temperature anisotropies in the cosmic microwave background (CMB).  
An exciting development
is the recent detection of the 3PCF of the cosmic shear field  by
Bernardeau, Mellier \& Van Waerbeke (2002a; also see Bernardeau, Van
Waerbeke \& Mellier 2002b), which provides new constraints on 
structure formation models beyond those provided by two-point
statistics. For the CMB, the bispectrum of the temperature anisotropies
is  a promising way of probing primordial non-Gaussian
perturbations (e.g., Verde et al. 2000; Komatsu et al. 2002). 
The 3PCF of the CMB polarization might open a new, complementary 
window for this in that it can separately measure the non-Gaussian
signals arising from primordial scalar, vector and tensor perturbations.

A spin-2 field has two components on the sky, hence its 3PCF in general
has $2^3 = 8$ components. One may ask: 
how many of these components are non-zero? Do all components carry useful 
information about the $B$- and $E$-modes of the spin-2 
field\footnote{A two-dimensional spin-2 field can be separated
into an $E$-mode derivable from a scalar potential and 
a pseudoscalar $B$-mode
(Kamionkowski, Kosowski \& Stebbins 1997; 
Zaldarriaga \& Seljak 1997; Hu \& White 1997 for the CMB polarization
and Stebbins 1996; Kamionkowski et al. 1998; Crittenden 
et al. 2001; Schneider et al. 2002 for the cosmic shear).}?
Recently, Schneider \& Lombardi (2002; hereafter SL02) and 
Zaldarriaga \& Scoccimarro (2002; hereafter ZS02) investigated
these questions. 
The purpose of this {\em Letter} is to clarify these issues
using geometrical arguments and measurements 
from ray-tracing simulations of the cosmic shear.  We will pay 
particular attention to the problem of how the 
eight 3PCFs are related to the $E$- and $B$-modes. 

\section{The 3PCF of Spin-2 Fields}

The two components of a spin-2 field depend on the choice of 
the coordinate system. Suppose we have the shear components, 
$\gamma_1$ and $\gamma_2$ on the sky, for given 
Cartesian coordinates\footnote{
We use the notations of weak lensing 
for the spin-2 field, $\gamma_i$; our discussion can be
applied to the CMB polarization if one replaces $\gamma_1$  and $\gamma_2$ 
with the Stokes parameters $Q$ and $U$, respectively.}.
A rotation of the coordinate system by $\varphi$ (in the anticlockwise
direction in our convention) transforms the shear fields as
$\gamma^\prime_1+i\gamma^\prime_2=(\gamma_1+i\gamma_2)e^{-i2\varphi}$.

For the two-point correlation function (2PCF), 
the problem of the coordinate dependence of the shear field 
has been well studied in the literature (e.g., 
Bartelmann \& Schneider 2001). 
For a given pair of points, 
$\bfm{X}_1$ and $\bfm{X}_2$,
separated by a fixed angle $\theta$, we can define two
components of the 2PCF which are invariant under coordinate rotations:
$
\xi_{+}(\theta)=\skaco{\gamma_+(\bfm{X}_1)\gamma_+(\bfm{X}_2)}$ and
$\xi_\times(\theta)=\skaco{\gamma_\times(\bfm{X}_1)
\gamma_\times(\bfm{X}_2)}$.
Here $\gamma_{+}(\bfm{X}_i)$ and $\gamma_{\times}(\bfm{X}_i)$ are 
the shear components defined by projecting the shear along the 
direction $\varphi$ connecting the two points: 
$\gamma_+(\bfm{X}_i)+i\gamma_\times(\bfm{X}_i)
=-\left[\gamma_1(\bfm{X}_i)+i\gamma_2(\bfm{X}_i)\right]e^{-2i\varphi}$. 
The other possible component,
$\skaco{\gamma_+\gamma_\times}$, vanishes because of parity 
invariance.

\begin{figure}[t]
\centerline{\epsfxsize=5.cm\epsffile{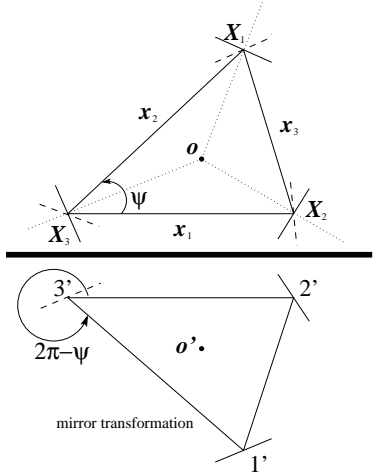}}
\caption{{\em Upper}: Definitions of the triangle variables used for the 3PCF. 
The interior angle
between $\bfm{x}_1$ and $\bfm{x}_2$ is defined as $\psi$ and its
 positive direction is shown by the arrow. 
Note that the solid and dashed curves at each vertex
show the positive directions for $\gamma_{+}$ and $\gamma_\times$.
{\em Lower:}  
The mirror transformation with respect to 
the side vector $\bfm{x}_1$ for $\zeta_{++\times}$. 
The vertices are transformed as $1\rightarrow 1'$ and so on. Note that
the $+$ components remain unchanged in sign, but the sign of the $\times$
component is flipped. 
\label{fig:triang}}
\end{figure}

SL02 and ZS02 investigated possible eight components of the 
shear 3PCF based on the $+/\times$ decomposition of the shear field. 
We will closely follow these authors in setting up the
geometry of the 3PCF. 
In contrast to the 2PCF, there is no unique choice 
of a reference direction to define the $+/\times$ decompositions
for the shear fields at the vertices of a triangle. 
The `center of mass', $\bfm{o}$, of a triangle 
is one possible fiducial choice:
$\bfm{o}\equiv (1/3)\sum_{i=1}^3\bfm{X}_i$. 
Figure \ref{fig:triang} shows a sketch of the triangle we use to 
define the shear 3PCFs. The solid and dashed lines at each vertex show
the positive directions of the $+$ and $\times$ components,
respectively. We also define the interior angle $\psi$ between
$\bfm{x}_1$ and $\bfm{x}_2$. 
The projection operator to compute the $+$ or $\times$ 
components at each vertex is written as:
\begin{eqnarray}
\bfm{P}_+(\bfm{X}_i)
&=&-(\theta_{i1}^2-\theta_{i2}^2,2\theta_{i1}\theta_{i2})/\theta_i^2,\nonumber\\
\bfm{P}_\times(\bfm{X}_i)&=&
-(-2\theta_{i1}\theta_{2i},\theta_{i1}^2-\theta_{i2}^2)/\theta_i^2,
\end{eqnarray}
where \hbox{\boldmath $\theta$}$_i\equiv \bfm{X}_i-\bfm{o}$.  
Using these projections, we obtain at each vertex 
$\gamma_{+} = {P}_{+1} \gamma_1 + 
{P}_{+2} \gamma_2$ and $\gamma_{\times} = {P}_{\times 1} \gamma_1 + 
{P}_{\times 2} \gamma_2$. These transform under parity as 
 $\gamma_{+} \rightarrow \gamma_{+} $ and $\gamma_{\times} 
\rightarrow - \gamma_{\times} $. 
In terms of $\gamma_+$ and $\gamma_\times$ we 
define the eight components of 
the shear 3PCF for a given triangle as a function of 
$x_1$,  $x_2$ and $\psi$: 
\begin{equation}
\zeta_{\mu\nu\tau}(x_1,x_2,\psi)
\equiv\skaco{\gamma_\mu(\bfm{X}_1)
\gamma_\nu(\bfm{X}_2)\gamma_\tau(\bfm{X}_3)},
\end{equation}
where $\mu, \nu, \tau= +$ or $\times$ and $\skaco{\cdots}$ denotes 
the ensemble average. The 3PCF defined above is invariant under
triangle rotations with respect to the center $\bfm{o}$ 
as pointed out in SL02. 
Hence it is fully characterized by the three variables 
$x_1, x_2, \psi$, just like the 3PCF of a scalar field (note that
if we had used the Cartesian components of the shear, we would need
four variables to specify it as it is not invariant under rotation). 

For a pure  $E$-mode,  
based on the properties of $\gamma_{+}$ and $\gamma_{\times}$
under parity transformations described above, 
we divide the eight  shear 3PCFs into two groups: 
\begin{eqnarray}
\mbox{Parity-even functions:  } & \zeta_{+++},\zeta_{+\times\times},
\zeta_{\times+\times},\zeta_{\times\times+}\nonumber \\
\mbox{Parity-odd functions:  } & \zeta_{\times\times\times},
\zeta_{\times++},
\zeta_{+\times+},\zeta_{++\times}.
\label{eqn:3pt}
\end{eqnarray}
Weak lensing only produces the $E$-mode,  
while source galaxy clustering, 
intrinsic alignments and observational systematics 
induce both $E$ and $B-$modes in general 
(Crittenden et al. 2001; Schneider et al. 2002)\footnote{Multiple 
lensing deflections generally induce
 the $B$-mode, but Jain, Seljak \& White (2000) showed
that the induced $B$-mode is much smaller than the lensing $E$-mode
using ray-tracing simulations.}.
For the CMB polarization, although primordial scalar perturbations
generate only the $E$-mode, vector and 
tensor perturbations can induce both modes (e.g., Kamionkowski et
al. 1997).  


A question posed by SL02 and ZS02 is whether 
all eight 3PCFs could have contributions from lensing.
We examine this question with geometrical considerations 
and show results from simulations in the next section. 
The parity transformation for a triangle can be taken to be a
mirror transformation as sketched in Figure \ref{fig:triang}, which
shows the parity transformation with respect to the side 
vector $\bfm{x}_1$ as one example. 
It corresponds to the change 
$\psi\rightarrow   2\pi-\psi$ in our parameters. 
Figure \ref{fig:triang} illustrates the
transformation for $\zeta_{++\times}$. From statistical homogeneity and
symmetry, the  amplitude
of the 3PCF depends only on the distances between the center and each
vertex. Hence, the absolute amplitudes of $\zeta_{++\times}$
for the two triangles shown should be the same. But the sign of
$\gamma_\times$ at the vertex $3'$ changes under this parity transformation. 
For the eight 3PCFs in equation (\ref{eqn:3pt}), 
we can say in general: 
\begin{eqnarray}
&& \mbox{Parity-even: }
\zeta_{\mu\nu\tau}(x_1,x_2,\psi)=\zeta_{\mu\nu\tau}(x_1,x_2,2\pi-\psi),
\nonumber\\
&& \mbox{Parity-odd: } \ 
\zeta_{\mu\nu\tau}(x_1,x_2,\psi)=-\zeta_{\mu\nu\tau}(x_1,x_2,2\pi-\psi).
\label{eqn:mirror}
\end{eqnarray}

Next, we consider an isosceles triangle with $x_1=x_2$. 
In this case, the $\gamma_\times$ at vertex $3$ and $3'$ in 
Figure \ref{fig:triang} are statistically identical (viewed from 
the center of the triangle, they should have equal contributions
when averaged over a matter distribution). 
We thus have the additional symmetries for the two parity-odd 3PCFs: 
\begin{eqnarray}
\zeta_{++\times}(x_1,x_1,\psi)&=&\zeta_{++\times}(x_1,x_1,2\pi-\psi),
\nonumber\\
\zeta_{\times\times\times}(x_1,x_1,\psi)
&=&\zeta_{\times\times\times}(x_1,x_1,2\pi-\psi).
\label{eqn:sym}
\end{eqnarray}
The properties described in 
equations (\ref{eqn:mirror}) and (\ref{eqn:sym}) lead to 
\begin{equation}
\mbox{Isosceles Triangles: } \  \ 
\zeta_{++\times}=0, 
\hspace{1em} \zeta_{\times\times\times}=0.
\label{eqn:parityinv}
\end{equation}
Note that this argument does not lead to the vanishing
of the other two parity-odd functions, in which the 
$\times$ component is at a vertex bounded by unequal sides. 
For equilateral triangles however all four parity-odd functions
vanish: 
\begin{equation}
{\rm Equilateral \ Triangles: } \ \ 
\zeta_{\rm parity-odd}=0. 
\label{eqn:equil}
\end{equation}
We argue that for a generic triangle with all sides unequal, 
the differences in side lengths
break the symmetry that is expressed 
in equation (\ref{eqn:sym}). In other words, the
fact that gravitation has a scale dependence allows parity-odd
3PCFs to be non-zero for a triangle with unequal sides. 
These conclusions are consistent with those of SL02. 

\section{Predictions from ray tracing simulations}

To demonstrate the properties of the shear 3PCFs more precisely,
we employ ray-tracing simulations of the cosmic shear
performed by Jain et al. (2000). We use the SCDM model
($\Omega_{\rm m0}=1$, $h=0.5$ and $\sigma_8=0.6$), with source redshift
$z_s=1$ and area  $\Omega_{\rm sky}=7.68$ degree$^2$ 
(see Jain et al. 2000 for more details). 
We followed Barriga \& Gazta\~naga (2002) for the algorithm to 
calculate the 3PCF. 
Lists of neighbors are used to find the three vertices from the
cell-based data as well as to calculate the projection operators for each
vertex. The error bars shown in the following figures are computed from 
9 different realizations (see Takada \& Jain 2002c for more details). 

\begin{figure}[t]
\centerline{\epsfxsize=8.5cm\epsffile{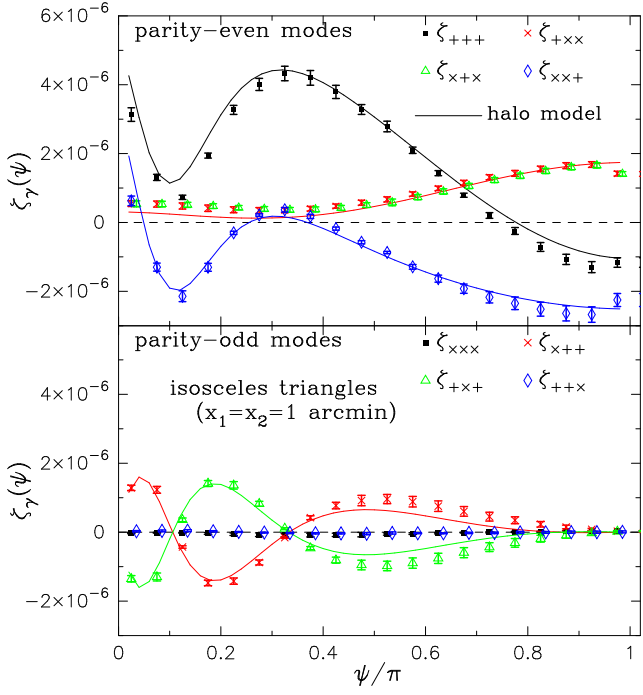}}
\caption{Ray-tracing simulation results of the shear 3PCF 
for isosceles triangles. The two side lengths are $x_1=x_2=0.\! \!'97$. 
The upper panel shows 
the four parity-even functions, while the lower panel shows the 
four parity-odd functions. For illustrative purpose, 
the results for $\zeta_{\times+\times}$ and $\zeta_{++\times}$ are 
slightly shifted horizontally. The solid curves
show analytical calculations of the shear 3PCFs based on the
halo model developed in Takada \& Jain (2002b). 
\label{fig:iso}}
\end{figure}
Figure \ref{fig:iso} shows the results for the shear 3PCFs 
for isosceles triangles against the interior angle $\psi$ between 
the two equal sides  $x_1=x_2=0.\! \! '97$ (see Figure \ref{fig:triang}). 
The shear 3PCF is computed from the simulated data by averaging the
estimator over all triplets with given triangle configuration. 
For the bin size we used
$\Delta x=0.\! \!'08$ for the side lengths and
$\Delta \psi=\pi/20$ for the interior angle.
The upper panel shows the four parity-even 3PCFs, while the lower panel
shows the parity-odd functions. As expected, 
$\zeta_{+++}$ appears to have the greatest 
lensing contributions, and, in particular,
peaks for equilateral triangles with $\psi=\pi/3$. 
The other three modes have smaller amplitude than $\zeta_{+++}$, and two of
them are the same because of symmetric triangle and shear configurations. 
Interestingly, the
features of these curves match the theoretical curves shown in Figure 3
of ZS02 ($\phi$ in their figures corresponds to $\pi-\psi$ in our figure). 
Their results were presented with arbitrary normalization of the y-axis,
so we can only compare the configuration dependence. 
Thus these small-scale complex features are basically captured
by the tangential shear patterns around a single halo.

From the lower panel in Figure \ref{fig:iso},
it is clear that two of the parity-odd 3PCFs
are non-zero, but the others 
vanish, as stated in equation (\ref{eqn:parityinv}). 
This result clarifies that the parity-odd 3PCFs in general
do carry lensing information and vanish only for special triangle
configurations. 
For isosceles triangles, the two vanishing parity-odd functions can be used 
to discriminate systematics from the $E$-mode. 
For $\psi=\pi/3$, the triangle is equilateral, and 
all the parity-odd 3PCFs vanish as stated in equation 
(\ref{eqn:equil}). 

We have verified that the simulation results are in agreement
with the 1-halo term predictions of the halo model. The solid curves
in Figure \ref{fig:iso} show the halo model predictions that follow
the real space halo formalism developed in Takada \& Jain (2002b). 
The calculations are based on adapting equation (52) from this paper
by replacing $\kappa_m$ with the relevant shear component for a halo
of given mass. A detailed comparison of analytical and simulation
results for different models will be presented in Takada \& Jain (2002c). 

\begin{figure}[t]
\centerline{\epsfxsize=8.5cm\epsffile{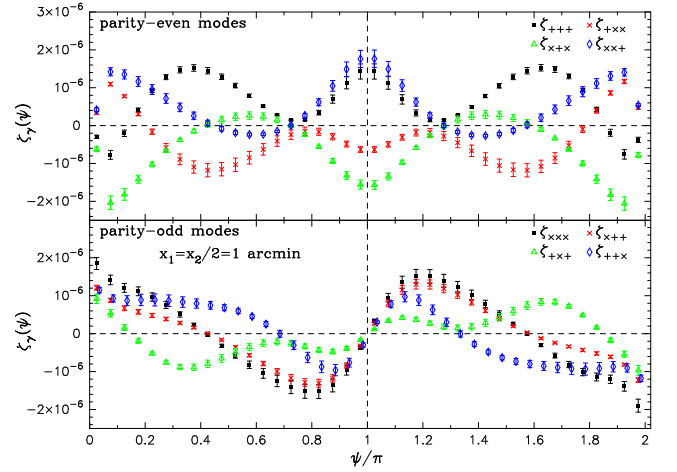}}
\caption{The results for a general triangle with $x_1=x_2/2=0.\! \!'97$,
as in Figure \ref{fig:iso}. 
\label{fig:general}}
\end{figure}

The features shown above for isosceles triangles
are explored for a triangle with three unequal sides in Figure 
\ref{fig:general}, in which $x_1=x_2/2=0.\! \! '97$. 
This figure is plotted over the range 
$\psi=[0,2\pi]$ to explicitly show 
the parity-even (upper panel) 
and  odd (lower panel) properties of the shear 3PCFs.
Note that the parity transformation is equivalent to the
change $\psi\rightarrow 2\pi-\psi$ in our parameters. 
As noted in ZS02, the  amplitude of the 3PCF is suppressed 
for elongated triangles because of cancellations in the signals by the
vector-like property of the shear. The figure
reveals that all the parity-odd 3PCFs
are non-zero and in fact all the eight 3PCFs
have roughly comparable amplitude. Note that
$\zeta_{\times\times\times}$ and $\zeta_{+\times+}$  vanish
at $\psi/\pi=0.42$, as the triangle becomes   
an isosceles triangle with $x_2=x_3$, and therefore the 3PCFs
are consistent with the result in Figure \ref{fig:iso}. 

\begin{figure}[t]
\centerline{\epsfxsize=8.5cm\epsffile{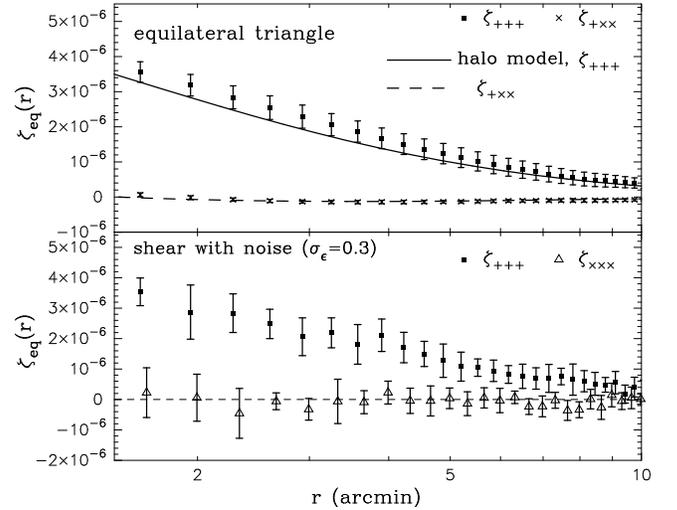}}
\caption{The upper panel shows 
the shear 3PCF for equilateral triangle as a function of the 
side length. Ray tracing measurements and halo model calculations
are shown for the two independent, non-zero 3PCFs, $\zeta_{+++}$ and
$\zeta_{\times++}$. The lower panel shows 
$\zeta_{+++}$ and
$\zeta_{\times\times\times}$ from simulated maps that include noise 
due to intrinsic ellipticities with rms $\sigma_{\epsilon}=0.3$. 
The bin size is $\Delta r=0'\!\!.325$.  
\label{fig:equil}}
\end{figure}

By rotating the simulated shear field at each position by $45$ degrees,
 we can investigate the 3PCF of a pure $B$-mode map (Kaiser 1992). Since 
this procedure transforms $\gamma_{+}
\rightarrow \gamma_{\times}$ and $\gamma_{\times}\rightarrow -
\gamma_+$,  
we find $\zeta^E_{+++}\rightarrow 
\zeta^{B}_{\times\times\times}$ and so on.  
Hence,  $\zeta^B_{\times\times\times}$ 
carries most of the $B$-mode signal for triangles that are close
to equilateral. 
Further the symmetry properties of the 3PCFs 
under $\psi \rightarrow 2\pi - \psi$ (shown in Figure 3) 
are reversed for the $B$-mode; 
for example, $\zeta^B_{\times\times\times}(\psi)=
\zeta^{B}_{\times\times\times}(2\pi-\psi)$. 
Therefore, we should keep in mind that for
a general spin-2 field which includes both $E$ and $B$ modes, 
the property $\zeta(\psi)=\pm \zeta(2\pi-\psi)$ can no longer be expected. 
The measurements of $\zeta$ thus has to be done for the full range
$\psi=[0,2\pi]$ (unlike the case for the 3PCF of 
a scalar quantity like the density field or the CMB temperature 
fluctuations field). 

The upper panel of 
Figure \ref{fig:equil} shows the results for the shear 
3PCF for equilateral triangles against the side length.
For the measurements from simulated data, 
we used triangles with
each side length in the range $[r-\Delta r,r+\Delta r]$,
with bin size $\Delta r =0'\!\!.325$ kept fixed. 
Since all the parity-odd 3PCFs vanish for 
an equilateral triangle, the results for 
$\zeta_{+++}$ and $\zeta_{+\times\times}
(=\zeta_{\times+\times}=\zeta_{+\times\times})$ 
are shown. 
We also show the halo model predictions for
the 1-halo terms of  $\zeta_{+++}$ and $\zeta_{+\times\times}$
by the solid and dashed curves, respectively.
It is again apparent 
that the halo model predictions are in good agreement with the 
simulation results. 

To estimate the signal-to-noise ($S/N$) for measuring 
the shear 3PCFs from an actual survey,
we need to account for the noise arising from the large intrinsic
ellipticities of source galaxies.
The lower panel of Figure \ref{fig:equil} shows 
the 3PCFs measured from simulated shear maps including 
the noise contamination. 
We assume that the intrinsic ellipticities have random
orientations with a Gaussian distribution for the amplitudes
with rms $\sigma_\epsilon=0.3$. 
The galaxies are taken to be randomly distributed with number density
$n_{\rm gal}\approx 38$ arcmin$^{-2}$. 
One can see that $\zeta_{+++}$ 
can be detected from a survey with area of order 10 square degrees, 
provided the errors are dominated by statistical errors. 
The other 3PCFs are more difficult to measure for equilateral 
triangles. 
Here we also show $\zeta_{\times\times\times}$ which is expected to
be zero for an $E$-mode signal, so the error bars on it show the
accuracy with which the $B$-mode can be constrained. 
These results in part verify the detection of the shear 3PCF
by Bernardeau et al. (2002a) from the cosmic shear survey 
of Van Waerbeke et al. (2001), although they used a different
estimator of the shear 3PCF. 
The $S/N$ for one particular triangle configuration with size $r$ 
 scales roughly as $S/N\propto 
\zeta_\gamma(\sigma_\epsilon/\sqrt{n_{\rm gal}})^{-3}
\Omega_{\rm survey}^{1/2}r^2 (\Delta \ln r)^{3/2}$, 
where $\Omega_{\rm survey}$ is the survey area. 
By combining information from triangles with different configurations
and from the different 3PCFs, we can improve the $S/N$ 
(however, 
to interpret the measured 3PCFs one must
estimate how correlated neighboring bins are, as discussed below). 
Note further that we have used the SCDM
model with $\zeta_{+++}=2-4\times 10^{-6}$ for
$r=1-5'$;  the $S/N$ for the
concordance $\Lambda$CDM model is lower because the signal is smaller, 
e.g.  $\zeta_{+++}=4-8\times 10^{-7}$ (Takada \& Jain 2002c). 
This rough analysis leads us to conclude that future surveys with 
area $\gsim 100$
degree$^2$ should provide measurements of the 3PCFs with $S/N$
well over 10 in each bin if systematic errors are eliminated. 

\section{Discussion}

We have investigated the 3PCF of the cosmic shear field. 
We measured the eight components of the shear 3PCF
from ray-tracing simulations and checked our results
using analytical calculations based on the halo model. 
Figures 2 and 3 show that in general all eight components are non-zero,   
have comparable amplitude, and have a complex configuration dependence. 
These results verify the analysis of SL02 who analyzed the 
3PCFs based on their transformations under parity and rotations. 

The 3PCFs show non-Gaussian signatures induced by non-linear 
gravitational clustering.  In constraining 
models from measurements of the shear 3PCFs, 
it will be necessary to develop an optimal way to combine
the eight 3PCFs. To do this, we will need explicit relations
showing how the shear 3PCFs are related to the lensing $E$-mode,
in analogy with two-point
statistics (Crittenden et al. 2001; Schneider et al. 1998, 2002).
This problem is more easily formulated in Fourier space. 
If we use the Cartesian components of the shear to define the 3PCFs, 
then  simple relations exist between the bispectra of the shear 3PCFs, 
$\tilde{\zeta}_{ijk}$ and the bispectrum of the convergence, $B_\kappa$: 
\begin{equation}
\tilde{\zeta}_{ijk}(\bfm{l}_1,\bfm{l}_2,\bfm{l}_3) = 
B_{\kappa}(\bfm{l}_1,\bfm{l}_2,\bfm{l}_3)
u_i(\bfm{l}_1) u_j(\bfm{l}_2) u_k(\bfm{l}_3) ,
\end{equation}
where $i, j, k = 1,2$ and $u_i(\bfm{l}) = (\rm{cos} 2\varphi_{\bfm{l}}, 
\rm{sin} 2\varphi_{\bfm{l}})$ for the $E$-mode. 
This equation shows that the shear
3PCFs can be related to a single 3PCF for the $E$-mode field. 
For example we can obtain the convergence bispectrum using the
eight measured shear bispectra as:
\begin{equation}
B_\kappa(\bfm{l}_1,\bfm{l}_2,\bfm{l}_3) = 
\sum_{i,j,k} u_i(\bfm{l}_1) u_j(\bfm{l}_2) u_k(\bfm{l}_3) 
\tilde{\zeta}_{ijk}(\bfm{l}_1,\bfm{l}_2,\bfm{l}_3) .
\end{equation}
However in real space it is still unclear how best to use the 
eight 3PCFs measured over a limited range of scales to get an optimal
estimator of the $E$-mode 3PCF. 
Another question we have not considered is how strongly
the shear 3PCFs are correlated for different triangle configurations, 
since lensing fields at arcminute scales are highly non-Gaussian 
(e.g. Takada \& Jain 2002a). 

Realistic data has noise which includes both $E/B$ modes, as would
contributions from intrinsic ellipticity correlations, nonlinear
lensing effects and systematic errors. 
We have shown that two or all four parity-odd 3PCFs vanish for 
isosceles or equilateral triangles, respectively (see Figures 2 and 4).  
For general triangles, Figure 3 shows that the 3PCFs of $E$-mode shear fields
have specific symmetry properties under the change
$\psi \rightarrow 2\pi - \psi$, where $\psi$ is the interior angle
of the triangle shown in Figure 1.  
These properties can be used to find
$B$-mode contributions (whose symmetry properties are reversed) as a function
of scale and configuration. Thus the origin of the $B$-mode contribution
can be identified more precisely than is possible with two-point statistics. 
An example of an explicit test for $B$ modes is to measure combinations 
such as $\zeta_{++\times}(x_1,x_2,\psi) + 
\zeta_{++\times}(x_1,x_2,2\pi-\psi)$. This should be zero 
for all the functions that are parity-odd for a pure $E$ mode, as given
in equation (\ref{eqn:3pt}). 

The results presented above can also be applied to 
the $Q$ and $U$ Stokes parameters of the CMB polarization.
The results we have 
shown imply that the eight 3PCFs constructed from combinations of 
$Q/U$, e.g. $\skaco{UQQ}$, have non-vanishing 
signals if the CMB polarization field is non-Gaussian.
The 3PCFs have the advantage that it can discriminate the non-Gaussianity
arising from the $E$- and $B$-modes.
However, since the polarization 
signal is small compared to the CMB temperature fluctuations, 
this will be a great challenge. 

\acknowledgments
We are grateful to G. Bernstein, L. Hui, M. Jarvis, P. Schneider, 
R. Scoccimarro, A. Stebbins, A. Szalay and M. Tegmark for helpful discussions. 
This work is supported by NASA grants NAG5-10923, NAG5-10924 
and a Keck foundation grant.


\begin{thebibliography}{}

\bibitem[Barriga \& Gazta\~naga 2002]{Barriga02}
  Barriga, J., \& Gazta\~naga, E. 2002, \mnras, 333, 443

\bibitem[Bartelmann \& Schneider 2001]{Bart01}
  Bartelmann, M., \& Schneider, P. 2001, Phys.~Rep. 340, 291

\bibitem[]{BMvW02}
  Bernardeau, F., Mellier, Y., \& 
Van Waerbeke, L. 2002a, \aap, 389, L28

\bibitem[]{BvWM02}
  Bernardeau, F., Van Waerbeke, \&
L., Mellier, Y. 2002b, astro-ph/0201029

\bibitem[]{Critt01}
  Crittenden, R.~ G., Natarajan, P., Pen, U.-L., \& Theuns, T. 2001, 
 \apj, 559, 552

\bibitem[]{HW97}
 Hu, W., \& White, M. 1997, \prd, 56, 596

\bibitem[]{JSW00}
 Jain, B., Seljak, U., \& White, S.~D.~M. 2000, ApJ, 530, 547

\bibitem[]{Kaiser92}
 Kaiser, N. 1992, \apj, 388, 272

\bibitem[]{Kam97}
 Kamionkowski, M., Kosowsky, A., \& Stebbins, A. 1997, \prd, 55, 7368

\bibitem[]{Kam98}
 Kamionkowski, M., Babul, A., Cress, C.~ M., \& Refregier, A. 1998,
\mnras, 301, 1064

\bibitem[]{Komatsu02}
 Komatsu, E., Wandelt, B.~ D., Spargel, D., Banday, A.~ J., 
\& G\'orski, K.~ M. 2002, \apj, 566, 19

\bibitem[]{SL02}
Schneider, P., \& Lombardi, M. 2002, astro-ph/0207454 (SL02)

\bibitem[]{Sch98}
Schneider, P., Van Waerbeke, Jain, B., \& Kruse, G. 1998, \mnras,
296, 873

\bibitem[]{Sch02}
Schneider, P., Van Waerbeke, L., Kilbinger, M., \& Mellier, Y. 2002, 
astro-ph/0206182

\bibitem[]{Stebbins02}
Stebbins, A., 1996, astro-ph/9609149 

\bibitem[]{TJ02a}
Takada, M., \& Jain, B. 2002a, \mnras, in press, astro-ph/0205055 

\bibitem[]{TJ02b}
Takada, M., \& Jain, B. 2002b, submitted to \mnras, astro-ph/0209167 

\bibitem[]{TJ02c}
Takada, M., \& Jain, B. 2002c, in preparation

\bibitem[]{Ludvic01}
Van Waerbeke et al. 2001, \aap, 374, 757

\bibitem[]{Verde00}
 Verde, L., Wang, L., Heavens, A.~ F., \& 
Kamionkowski, M. 2000, \mnras, 313, 141 

\bibitem[]{ZS02}
Zaldarriaga, M., \& Scoccimarro, R. 2002, astro-ph/0208075 (ZS02)

\bibitem[]{Zald97}
Zaldarriaga, M., \& Seljak, U. 1997, \prd, 55, 1830

\end{thebibliography}
\end{document}